\documentstyle[aps,twocolumn,prb,psfig]{revtex}
\begin{document}
\draft
\title{
Interplay of the CE-type charge ordering and the A-type spin ordering\\
in a half-doped bilayer manganite La$_{1}$Sr$_{2}$Mn$_{2}$O$_{7}$
}
\author{M. Kubota,$^{1}$ H. Yoshizawa,$^{1}$ Y. Moritomo,$^{2}$ H. Fujioka,$^{3}$ K. Hirota,$^{3}$ and Y. Endoh$^{3}$}
\address{$^{1}$Neutron Scattering Laboratory, I. S. S. P., University of
Tokyo, Tokai, Ibaraki, 319-1106, Japan}
\address{$^{2}$ PRESTO, Center for Integrated Research in Science and Engineering (CIRSE)\\
Nagoya University, Nagoya, 464-8601, Japan}
\address{$^{3}$CREST, Department of Physics, Tohoku University, Aoba-ku, Sendai, 980-8578, Japan}
\date{\today}

\twocolumn[\hsize\textwidth\columnwidth\hsize\csname @twocolumnfalse\endcsname
\maketitle

\begin{abstract}
We demonstrate that the half-doped bilayer manganite La$_{1}$Sr$_{2}$Mn$_{2}$O$_{7}$ exhibits CE-type charge-ordered and spin-ordered states below $T_{\rm N, CO}^{\rm A}$=210 K and below $T_{\rm N}^{\rm CE} \sim145$ K, respectively.  However, the volume fraction of the CE-type ordering is relatively small, and the system is dominated by the A-type spin ordering.  The coexistence of the two types of ordering is essential to understand its transport properties, and we argue that it can be viewed as an effective phase separation between the metallic $d(x^{2}-y^{2})$ orbital ordering and the charge-localized $d(3x^{2}-r^{2})$/$d(3y^{2}-r^{2})$ orbital ordering.

\end{abstract}
\pacs{75.25.+z, 71.27.+a, 71.30.+h}
]

It is well known that a Mn-based Ruddlesden-Popper compound (La,Sr)$_{n+1}$Mn$_n$O$_{3n+1}$ has a unique layered-type structure in which every $n-$MnO$_2$ layers are separated by an additional (La$_{1-x}$Sr$_{x}$)$_2$O$_2$ blocking bilayer along the $c$ axis.\cite{mor96}  One of the attractive features of this family of compounds is that, by changing the number of MnO$_2$ layers ($n$), effective dimensionality can be artificially controlled.  The change of effective dimensionality have the strong influence on the magnetism as well as transport properties.\cite{mor96}   For instance, the 3d $n=\infty$ system La$_{1-x}$Sr$_{x}$MnO$_{3}$ is a ferromagnetic metal for $0.17 < x \lesssim 0.6$ due to the double-exchange mechanism,\cite{uru95} whereas the 2d $n=1$ compound La$_{1+x}$Sr$_{1-x}$MnO$_{4}$ is a good insulator for $0 \leq  x \lesssim 0.7$ and exhibits spin glass behavior for a wide range of the hole concentration.\cite{Y.Moritomo95}  For the $n=2$ bilayer La$_{2-2x}$Sr$_{1+2x}$Mn$_{2}$O$_{7}$ system, we have very recently reported that a canted A-type antiferromagnetic (AFM) structure is stabilized for the hole concentration of $0.40\lesssim x < 0.48$ (Ref. \onlinecite{hir98}).

One of the important aspects of perovskite manganites is the charge ordering which is frequently observed at a commensurate hole concentration.\cite{cheong}  In the vicinity of $x=1/2$, a number of the 3d $n=\infty$ manganite systems including Pr$_{1/2}$Ca$_{1/2}$MnO$_{3}$ (Ref. \onlinecite{jirak,yoshi95}) shows the so-called CE-type charge and spin ordering,  in which Mn$^{3+}$ and Mn$^{4+}$ ions are aligned alternately.\cite{wollan}  Neutron and X-ray diffraction experiments revealed that this unique CE-type charge/spin ordering is also realized in the $n=1$ La$_{1/2}$Sr$_{3/2}$MnO$_{4}$ system.\cite{J.Sternlieb96,mur98}  Very recently, an electron diffraction measurement suggested that the CE-type charge ordering is also formed in the $n=2$ La$_{1}$Sr$_{2}$Mn$_{2}$O$_{7}$ system.\cite{J.Q.Li98}

\begin{figure}[htb]
\centering
\psfig{file=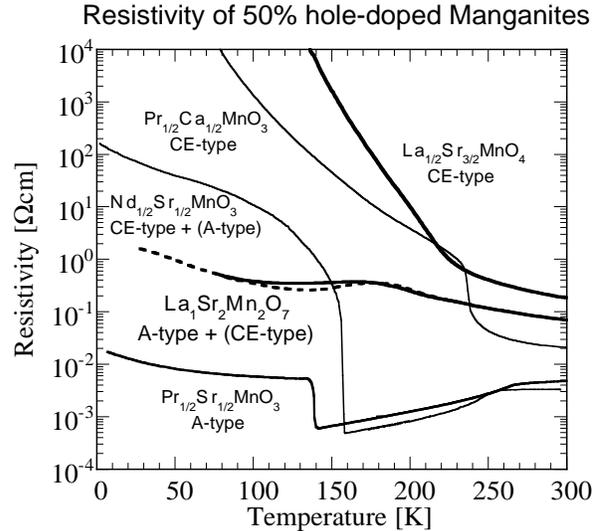,width=8cm}
\caption{Resistivity of 50 \% hole-doped 2d and 3d manganite samples: Pr$_{1/2}$Ca$_{1/2}$MnO$_{3}$ from Ref.\protect\onlinecite{tom95a},  Nd$_{1/2}$Sr$_{1/2}$MnO$_{3}$ from Ref.\protect\onlinecite{Kuwahara95},  Pr$_{1/2}$Sr$_{1/2}$MnO$_{3}$ from Ref.\protect\onlinecite{tom95b}, the $ab$-plane resistivities in La$_{1}$Sr$_{2}$Mn$_{2}$O$_{7}$ and in La$_{1/2}$Sr$_{3/2}$MnO$_{4}$ by the present work.}
\label{RT-dep}
\end{figure}

To see the influence of the CE-type charge ordering, the typical examples of the resistivity of several manganites with $x=0.50$ are depicted in Fig. \ref{RT-dep}.  The resistivity of the 3d $n=\infty$ system Pr$_{1/2}$Ca$_{1/2}$MnO$_{3}$ (Ref. \onlinecite{tom95a}) and that of the 2d $n=1$ system La$_{1/2}$Sr$_{3/2}$MnO$_{4}$ are insulating over the whole temperature range, and they exhibit a sharp increase below $T_{\rm CO}$.  By contrast, the resistivity of Pr$_{1/2}$Sr$_{1/2}$MnO$_{3}$ is considerably small. \cite{tom95b,kaw97} This is because a relatively wide one-electron bandwidth $W$ in Pr$_{1/2}$Sr$_{1/2}$MnO$_{3}$ suppresses the CE-type charge ordering, and leads to the A-type AFM spin ordering with very small resistivity ($\rho\sim7\times10^{-3} \Omega$cm) below $T_{N} \sim 138$ K.  Interestingly, the $ab$-plane resistivity in La$_{1}$Sr$_{2}$Mn$_{2}$O$_{7}$ below $T_{\rm CO}$ shows a rather flat temperature dependence similar to that of Pr$_{1/2}$Sr$_{1/2}$MnO$_{3}$, although it is two order of magnitude larger ($\rho\sim5\times10^{-1} \Omega$cm) than that of Pr$_{1/2}$Sr$_{1/2}$MnO$_{3}$.

It would be of great interest to elucidate what sort of mechanism could possibly cause such a drastic difference of the resistivity among 50 \% hole-doped manganite samples, and why only a difference of one MnO$_{2}$ layer leads to totally different behavior in the resistivity between two good 2d systems $n=1$ La$_{1/2}$Sr$_{3/2}$MnO$_{4}$ and $n=2$ La$_{1}$Sr$_{2}$Mn$_{2}$O$_{7}$.  In order to obtain some insights on these issues, we have performed neutron diffraction studies on the high quality large single crystal sample of the 50 \% hole-doped bilayer system La$_{1}$Sr$_{2}$Mn$_{2}$O$_{7}$.

For the present neutron diffraction study, we have prepared high quality single crystal samples by floating zone method as reported elsewhere.\cite{hir98}  The sample quality was checked by powder x-ray diffraction measurements to confirm that all the crystals were in a single phase.  The resistivity was measured for several crystals including the pieces taken from the same single crystal used in the present neutron diffraction studies, and the results are in excellent accord with the preceding work.\cite{J.Q.Li98,mor99}

Neutron diffraction measurements were carried out on triple-axis spectrometers GPTAS, TOPAN, and HQR in the JRR-3M research reactor at the Japan Atomic Energy Research Institute.  The spectrometers were operated in their double axis mode, and the $(0 0 2)$ reflection of pyrolytic graphite (PG) was used to monochromate the neutron beam.  At GPTAS,  the measurements were carried out on both the $(h, h, l)$ and $(h, 0, l)$ reciprocal zones with an incident momentum of $k_i = 2.667$\AA$^{-1}$, together with PG filters to eliminate higher order contaminations, and with a typical combination of collimators of $20^{\prime}-80^{\prime}-40^{\prime}$.  The lattice constants of the sample at room temperature are $a=3.875$\AA, and $c=19.971$\AA, respectively.

\begin{figure}[htb]
\centering
\psfig{file=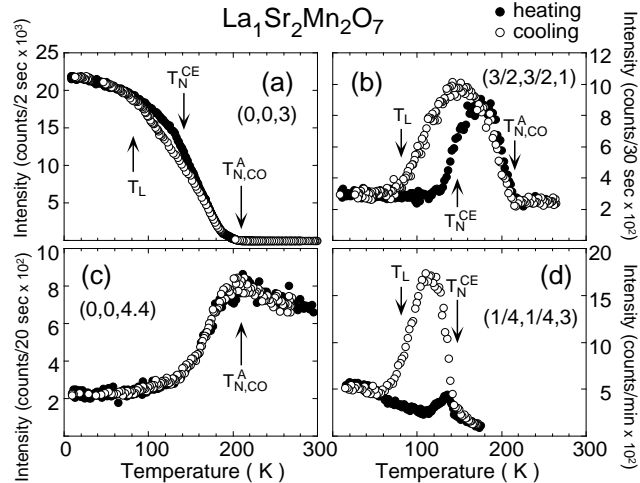,width=8.5cm}
\caption{Temperature dependence of selected positions for La$_{1}$Sr$_{2}$Mn$_2$O$_7$:  (a) A-type AFM Bragg, (b) CE-type charge order, (c) magnetic diffuse scattering, (d) CE-type AFM Bragg.}
\label{T-dep}
\end{figure}

We first describe the ordering process observed in La$_{1}$Sr$_{2}$Mn$_{2}$O$_7$.  Figure \ref{T-dep} shows the temperature dependences of Bragg and diffuse peaks in La$_{1}$Sr$_{2}$Mn$_{2}$O$_7$.  The most striking result of the bilayer system La$_{1}$Sr$_{2}$Mn$_{2}$O$_7$ is the coexistence of the A-type AFM spin ordering and the CE-type charge ordering.  The intensity at ${\bf Q}=(0,0,3)$ corresponds to the squared moment of the A-type AFM spin ordering, while the intensity at ${\bf Q}=(\frac{3}{2},\frac{3}{2},1)$ gives a measure of the order parameter of the CE-type charge ordering.  At $T_{\rm N, CO}^{\rm A}=210$ K, the A-type AFM spin ordering and the CE-type charge ordering are formed simultaneously, as seen in Fig. \ref{T-dep}(a) and (b).  With decreasing temperature, the intensity of ${\bf Q}=(0,0,3)$ increases monotonically, while that of ${\bf Q}=(\frac{3}{2},\frac{3}{2},1)$ reaches a maximum at $T_{\rm N}^{\rm CE} \sim 145$ K, and turns to decrease, and almost vanishes at $T_{\rm L} \sim 75$ K.  By contrast, the intensity of the CE-type spin ordering emerges at $T_{\rm N}^{\rm CE} \sim 145$ K, and exhibits a peak at $\sim 115$ K, and drastically decreases below $T_{\rm L}$ as shown in Fig. \ref{T-dep}(d).  For all known manganites such as Pr$_{1-x}$Ca$_{x}$MnO$_{3}$, La$_{1/2}$Sr$_{3/2}$MnO$_{4}$, and Nd$_{1/2}$Sr$_{1/2}$MnO$_{3}$, the CE-type charge/spin ordering is stable at low temperatures.\cite{yoshi95,J.Sternlieb96,Kuwahara95}  By contrast, the CE-type charge/spin-ordered state in La$_{1}$Sr$_{2}$Mn$_2$O$_7$ is unstable at low temperatures, and transforms to the metallic A-type AFM state.  It should be noted, however, that the substantial intensity for the CE-type charge/spin ordering is still discernible below $T_{\rm L} \sim 75$ K in Fig. \ref{T-dep}(b) and (d).

As was seen in the resistivity shown in Fig. \ref{RT-dep}, salient hysteresis was also observed in the temperature dependences of the A-type and CE-type charge/spin orderings in Fig. \ref{T-dep} for a wide temperature region of $T_{\rm L} \lesssim T \lesssim T_{\rm N, CO}^{\rm A}$.  Clearly, the hysteresis should be attributed to the onset and the suppression of the CE-type charge/spin ordering.  The temperature dependence of the magnetic diffuse scattering in Fig. \ref{T-dep}(c) observed at {\bf Q}= (0, 0, 4.4) demonstrates that the strong FM spin correlations have already been well developed within the MnO$_2$ layers from room temperature down to $T_{\rm N, CO}^{\rm A}$.

\begin{figure}[htb]
\centering
\psfig{file=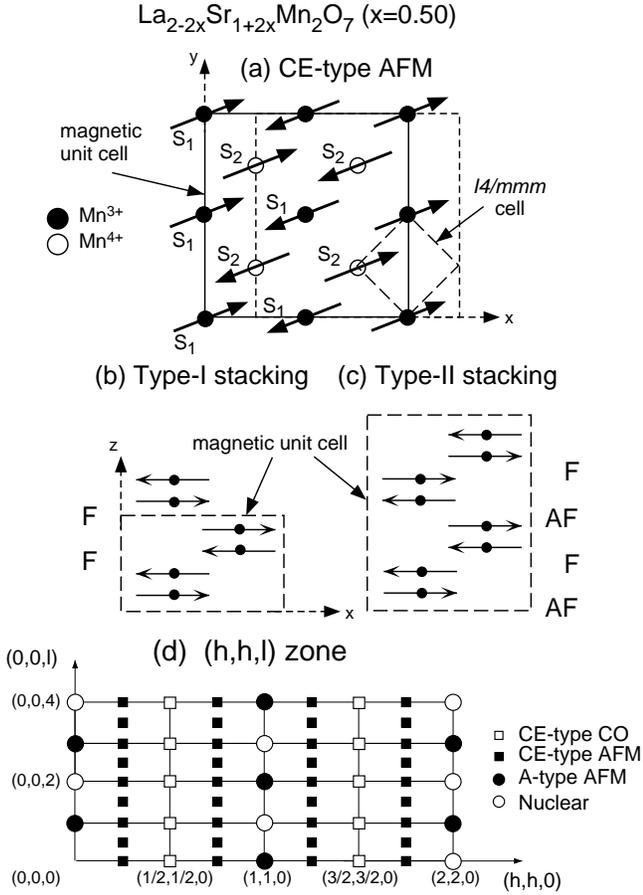,width=8.5cm}
\caption{ (a): CE-type spin arrangement in the $ab$ plane. The dashed square indicates the offset pattern at the $z=1/2$ layer by a displacement vector ${\bf d}=(\frac{1}{4},0,0)_{M}$. (b), (c): two stacking patterns along the $c$ axis. `F' and `AF' denote the ferromagnetic or antiferromagnetic coupling between adjacent bilayers, respectively. (See the text in details)  (d): $(h, h, l)$ reciprocal plane for La$_{1}$Sr$_{2}$Mn$_{2}$O$_{7}$.}
\label{recip}
\end{figure}

The observed structure for the CE-type charge and spin ordering in the bilayer system La$_{1}$Sr$_{2}$Mn$_{2}$O$_{7}$ is essentially the same with the known CE-type ordering except for the stacking along the $c$ axis.  The CE-type spin pattern in the $ab$ plane is depicted in Fig. \ref{recip}(a), and the observed Bragg spots on the $(h,h,l)$ reciprocal plane is illustrated in Fig. \ref{recip}(d).  The superlattice reflections due to the lattice distortion accompanied with the CE-type charge ordering are observed at $(h, h, l) = (\frac{2n+1}{2}, \frac{2n+1}{2}, m)$, while those for the CE-type AFM spin ordering are observed at $(h, h, l) = (\frac{2n+1}{4}, \frac{2n+1}{4}, l)$ with $n, m =$ integer, and $l =$ integer or half integer.

In the $(h, h, l)$ zone, however, all the magnetic reflections are originated from the Mn$^{3+}$ ions.  Since the reflections which are originated from the Mn$^{4+}$ sites in the CE-type spin ordering can be observed on the $(h, 0, l)$ zone, we have performed a survey of the $(h, 0, l)$ zone as well, and confirmed that the reflections by Mn$^{4+}$ sites are observed at the $(\frac{1}{2}, 0, l)$ positions with $l=$ integer, and half integer.

\begin{figure}[htb]
\centering
\psfig{file=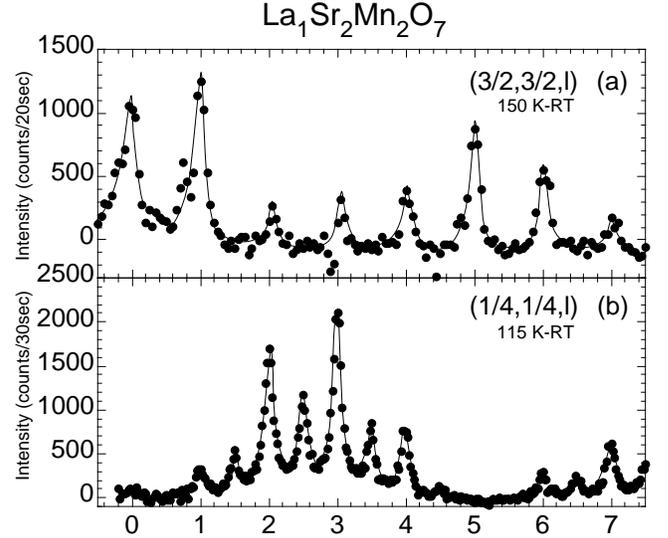,width=8.5cm}
\caption{(a) Profile of structural scattering observed along ${\bf q} = (\frac{3}{2}, \frac{3}{2}, l)$, and (b) profile of CE-type spin order observed along ${\bf q} = (\frac{1}{4}, \frac{1}{4}, l)$. For details, see the text.}
\label{spinprof}
\end{figure}

Figure \ref{spinprof} shows the typical profiles for the CE-type charge and spin ordering observed along the $l$ direction in the $(h, h, l)$ zone.  The measurements were carried out at 150 K and at 115 K, respectively, where the each scattering intensity exhibited the maxima, and the data measured at room temperature were subtracted as a back ground.\cite{BGsub98}  We confirmed that the peaks for the charge ordering on the $(\frac{3}{2}, \frac{3}{2}, l)$ line were stronger than those on the $(\frac{1}{2}, \frac{1}{2}, l)$ line, whereas the peaks for the CE-type AFM spin ordering on the $(\frac{3}{4}, \frac{3}{4}, l)$ line were weaker than those on the $(\frac{1}{4}, \frac{1}{4}, l)$ line, being consistent with the magnetic form factor.  We therefore identify that the former scattering is originated from the charge ordering, but the latter from the magnetic scattering.

Reflecting a bilayer structure of the $n=2$ La$_{1+2x}$Sr$_{2-2x}$Mn$_{2}$O$_{7}$ system, the Bragg intensities for both charge and spin ordering are modulated along the $l$ direction as,
\begin{eqnarray}
I(\frac{2n+1}{2}, \frac{2n+1}{2}, l) &\propto& 1+\cos 2\pi l z' \quad{\rm for \quad charge,}\\ 
I(\frac{2n+1}{4}, \frac{2n+1}{4}, l) &\propto& 1-\cos 2\pi l z' \quad{\rm for \quad spin,} 
\end{eqnarray}
where $z'$ is the Mn-Mn distance between MnO$_2$ layers within bilayers.  Note that,  due to the AFM coupling of two MnO$_2$ layers within bilayers, the sign of the cosine term for the magnetic Bragg reflections becomes minus in eq. (2). 

Although the peaks of the charge ordering are observed only at $l=$ integer, those of the spin ordering are observed both at $l=$ integer and half integer.  This difference is caused by the fact that there exist two kinds of coupling of bilayers along the $c$ axis as depicted in Fig. \ref{recip}(b) and (c).  From a straight forward analysis, we found that the peaks given by $l=$ integer correspond to the ferromagnetic stacking of bilayers along the $c$ axis ({\it Type-I stacking}) as indicated by `F-F' in Fig. \ref{recip}(b).  On the other hand, the peaks observed at $l=$ half integer correspond to a stacking sequence of `F-AF-F-AF' ({\it Type-II stacking}).  In addition, the profile shown in Fig. \ref{spinprof}(b) has a broad component beneath the Bragg reflections, which corresponds to the magnetic diffuse scattering and is again modulated as $I(l) \propto 1-\cos 2\pi l z'$.  This indicates that, in addition to the regular Type-I and Type-II stackings, there are some  bilayers which  are randomly stacked along the $c$ axis, reflecting a good 2d character of La$_{1}$Sr$_{2}$Mn$_{2}$O$_{7}$.

The analysis further revealed a shift of the CE-type spin patterns relative to the next bilayers.  By going to the adjacent bilayers, the CE-type spin arrangement has to shift by either ${\bf d}=(\pm\frac{1}{4},0,0)_{M}$, $(0,\pm\frac{1}{4},0)_{M}$ or their linear combination, reflecting the body-centered tetragonal structure of the crystal.  The case of ${\bf d}=(\frac{1}{4},0,0)_{M}$ in the units of the magnetic unit cell is illustrated by a dashed line in Fig. \ref{recip}(a), and our analysis yielded that the observed magnetic structure is consistent with ${\bf d}=(\pm\frac{1}{4},0,0)_{M}$ for both the Type-I and Type-II stackings.

Among the half-doped manganites with the CE-type charge/spin ordering, La$_{1}$Sr$_{2}$Mn$_2$O$_7$ is very unusual.  The preceding resistivity measurements and the electron diffraction study lead the speculation that La$_{1}$Sr$_{2}$Mn$_2$O$_7$ has an intermediate-temperature charge-ordered phase.\cite{J.Q.Li98}  We found that it is correct, but the CE-type charge ordering coexists with the A-type AFM ordering.  One can see from Fig. \ref{T-dep} that the scattering from the CE-type ordering is weak, and that the scattering is dominated by the A-type AFM ordering {\it even in the intermediate temperature phase} for $T_{\rm L} \lesssim T \lesssim T_{\rm N, CO}^{\rm A}$.  This observation is corroborated by the analysis of the observed magnetic moments.  We obtained the magnetic moments for the CE-type spin ordering at 115 K $\mu_{CE} = 0.70(5) \mu_B$/Mn for the Type-I stacking and $\mu_{CE} = 0.56(6) \mu_B$/Mn for the Type-II stacking, respectively, whereas the A-type moment was $\mu_{A} = 2.3(1) \mu_B$/Mn at 115 K.  Clearly the apparently small moment for the CE-type spin ordering is an artifact and is a strong indication of a small volume of the CE-type region.  By assuming that the magnetic moments in the CE-type and A-type spin ordering are approximately equal, we estimated that the volume fraction of the CE-type spin ordering is $\approx 18$\% at 115 K.

Examining the resistivity of the half-doped manganites in Fig. \ref{RT-dep}, one can notice that the resistivity has a clear correlation with the relative stability of the CE-type charge-ordered state against the metallic A-type AFM state.  Pr$_{1-x}$Ca$_{x}$MnO$_{3}$ and La$_{1/2}$Sr$_{3/2}$MnO$_{4}$ show no A-type AFM state, and they exhibit very large resistivity at low temperatures.  Unlike these CE-type charge-ordered systems, Nd$_{1/2}$Sr$_{1/2}$MnO$_{3}$ shows rather small resistivity of $\sim 10^{2} \Omega$cm, and it was recently discovered that this system exhibits a parasitic A-type AFM state in the CE-type charge-ordered phase.\cite{kawano98}  The resistivity of La$_{1}$Sr$_{2}$Mn$_2$O$_7$ is an order of $\sim 10^{0} \Omega$cm,\cite{mor99} and we have demonstrated in the present work that La$_{1}$Sr$_{2}$Mn$_2$O$_7$ is dominated by the A-type ordering both in the intermediate temperature phase for $T_{\rm L} \lesssim T \lesssim T_{\rm N, CO}^{\rm A}$ and at low temperatures with the remnant CE-type charge-ordered region.  Finally the metallic A-type AFM Pr$_{1/2}$Sr$_{1/2}$MnO$_{3}$ has the resistivity of $\sim 10^{-2} \Omega$cm.

This trend of the resistivity should be attributed to a metallic character of the A-type AFM spin ordering.\cite{kaw97}  Since spins are aligned ferromagnetically within MnO$_2$ planes in the A-type ordering, the double-exchange interaction is actuated, and yields the metallic in-plane conductivity.  Recently, the metallic conductivity of the A-type AFM state is reported in several heavily-doped manganite systems.\cite{kaw97,mor98}  Figure \ref{RT-dep} indicates that, as the fraction of the A-type state increases, the low temperature resistivity decreases.

It is intriguing to examine whether these results can be understood on analogy of phase separation, which is recently proposed for the less-doped manganite case.\cite{yunoki}  The fluctuation of charge density is essential in the phase separation model, and it is naturally built-in in a less-doped manganite system because the charge-density fluctuation may drive the system to split into a low hole-density insulating AFM region or into a high hole-density metallic FM region.\cite{yunoki}  For the 50 \% hole-doped systems, such charge fluctuation is unlikely.  On the other hand, the present results indicate that there can be two stable phases near 50 \% hole-doping due to the orbital and charge ordering.  In the A-type AFM state of La$_{1}$Sr$_{2}$Mn$_2$O$_7$, the $e_g$ electrons supposedly occupy the $d(x^{2}-y^{2})$ orbital,\cite{hir98} whereas in the CE-type charge-ordered state,  the $e_g$ electrons occupy $d(3x^{2}-r^{2})$ and $d(3y^{2}-r^{2})$ orbitals alternately. \cite{mur98}  The $d(x^{2}-y^{2})$ orbital is isotropic and charges are delocalized, while the $d(3x^{2}-r^{2})$ and $d(3y^{2}-r^{2})$ orbitals are accompanied with large Jahn-Teller distortions of MnO$_6$ octahedra, and clearly the charges are localized in the CE-type charge/spin ordering.  Combining the existing results on the half-doped manganites with our present observations, we suggest that, in general, the CE-type charge/spin ordering coexists with the A-type spin ordering at any desired ratio in nearly 50 \% hole-doped systems, and such a coexistence between two orbital-ordered states could be viewed as an effective phase separation.  We note that somewhat different case of microscopic phase separation is observed in La$_{1-x}$Ca$_{x}$MnO$_3$. \cite{mori}

In conclusion, we have demonstrated that the bilayer manganite La$_{1}$Sr$_{2}$Mn$_2$O$_7$ exhibits a CE-type charge/spin ordered state at intermediate temperature, but this state coexists with the dominant A-type antiferromagnetic state.  With decreasing temperature, the CE-type state is suppressed through a first order transition.  Strong hysteresis was observed for $T_{\rm L} \lesssim T \lesssim T_{\rm N,CO}^{\rm A}$.  Reflecting the dominant A-type orbital and spin ordering, however, the resistivity of La$_{1}$Sr$_{2}$Mn$_2$O$_7$ is extremely low, and the anomaly in the resistivity around the CE-type charge ordering is weak.  These results indicate the importance of the competition between the metallic A-type $d(x^{2}-y^{2})$ orbital ordering and the insulating CE-type $d(3x^{2}-r^{2})$/$d(3y^{2}-r^{2})$ orbital ordering near half-doped manganites.

We thank N. Furukawa, E. Daggoto and H. Kawano for valuable discussions, and Y. Tokura for providing numerical data in Ref. \onlinecite{tom95a,Kuwahara95,tom95b}.  This work was supported by a Grant-In-Aid for Scientific Research from the Ministry of Education, Science and Culture, Japan, and the work in Nagoya was supported by Precursory Research for Embryonic Science and Technology (PRESTO), Japan Science and Technology Cooperation (JST), Japan.


\begin{references}
\bibitem{BGsub98} From the hysteresis, we suspect the possible existence of the structural transition between $T_{\rm N}^{\rm CE} \sim 145$ K and $T_{\rm L} \sim 75$ K.  However, because there is no influence of the nuclear Bragg points on these lines and the change of the lattice constants is negligible, we substract the room temperature data as a back ground.
\end{references}
\end{document}